# A Novel Radial-Flux IPM Eddy-Current Coupler for Wind Generator Applications

Sajjad Mohammadi
*Electrical Engineering and Computer Science*
*MIT*
Cambridge, USA
sajjadm@mit.edu

Gholamreza Davarpanah
*Department of Electrical Engineering*
Amirkabir University of Technology
Tehran, Iran
ghr.davarpanah@aut.ac.ir

James L. Kirtley
*Electrical Engineering and Computer Science*
*MIT*
Cambridge, USA
kirtley@mit.edu

*Abstract*—A novel radial-flux eddy-current coupler with interior permanent magnets (IPM) is proposed, providing higher demagnetization tolerance, making it well-suited for applications with limited accessibility, such as offshore wind generation. Finite element analysis is employed in the design and derivation of coupler quantities. Finally, the coupler is prototyped to experimentally validate the design and the simulation results.

## I. Introduction

Eddy-current couplers [1]-[2], with increasing demand, have found applications in flywheel energy storage, high-speed levitated technologies, and precision robotic arms, addressing the need for non-contact torque transfer and minimized mechanical wear. Their use extends to wind turbines, offering inherent vibration filtering and overload protection [3]-[5]. In [6], PM synchronous couplers have been studied. Eddy-current couplers and other electric machines can be analyzed using finite element methods (FEM), a powerful tool, yet too slow for initial designs [7]-[9]. An alternative is analytical models developed using flux-tubes [10]-[13] and the solution of Laplace's and Poisson's equations [14]-[20] which fast yet accurate enough for design optimizations. Couplers with surface permanent magnets (SPM) have been studied in [21]-[24]. In [25]-[26], IPM eddy-current couplers are developed and analytically modeled.

The main contribution of this paper is to propose a novel radial-flux IPM eddy-current coupler with higher demagnetization tolerance compared to SPM topologies, where PMs are directly exposed to reaction fields and heat generated from induced currents. This, along with inherent vibration filtering, makes the proposed design suitable for wind generation applications. Magnetic fields, eddy currents, and torque are obtained using FEM, which is then verified by experimental results from a prototyped coupler.

## II. Proposed Topology

The topology and specifications of the proposed coupler are presented in Fig. 1 and Table I. As shown in Fig. 2, the rotation of one side of the coupler, connected to the prime mover, induces eddy currents in the conductive sheet (CS) due to the relative speed between the two sides. A torque is then developed from the interaction between these induced currents and the total air-gap magnetic field, which is the spatial sum of the field produced by the magnets and the reaction field originating from the eddy currents. Unlike SPM couplers, the PMs are not directly exposed to the reaction fields and heat generated by the induced eddy currents in the conductive sheet, which increases the device's demagnetization tolerance. The CS extends with an overhang to provide a return path for the induced currents.

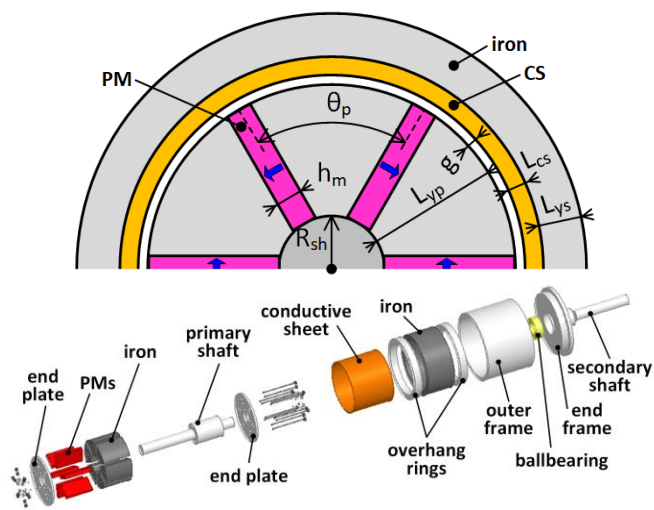

Fig. 1. Topology (top) and exploded view (bottom) of the proposed coupler

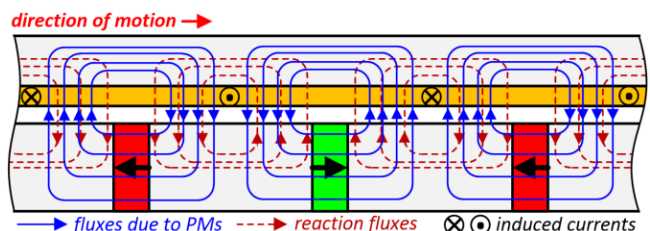

Fig. 2. PM field, induced eddy-currents and reaction fields

TABLE I. Specifications of the Coupler

| Parameter | Value | Parameter | Value |
|---|---|---|---|
| PM height, $h_m$ | 5 mm | Shaft radius, $R_{sh}$ | 15 mm |
| Air-gap length, $g$ | 0.5 mm | CS overhang, $H$ | 10 mm |
| CS thickness, $L_{cs}$ | 1 mm | Active axial length, $L$ | 40 mm |
| Inner yoke length, $L_{yp}$ | 20 mm | Number of PMs, $N_{pm}$ | 6 |
| Outer yoke length, $L_{ys}$ | 8 mm | NdFeB PM grade | N35 |

## III. Field Analysis and Experimental Study

Magnetic fields and eddy currents are obtained using 2-D and 3-D FEM. As shown in Fig. 3(a), each flux loop in the IPM

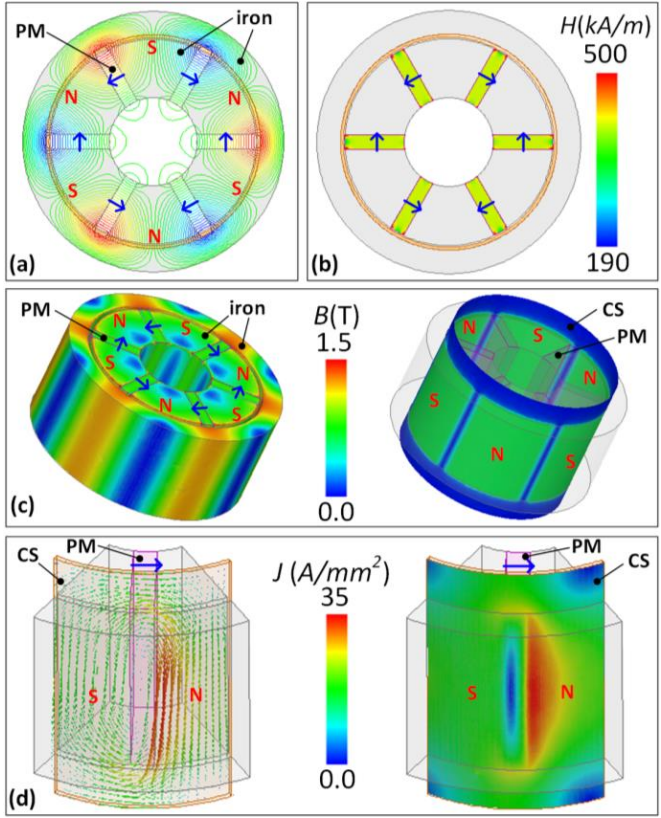

Fig. 3. FEM: (a) flux lines, (b) field intensity within PMs, (c) flux density on the surface of iron, PMs and CS, and (d) eddy-current vectors and distribution

topology is associated with one magnet. There is also leakage flux through the shaft and top and bottom surfaces, which is an inherent drawback of IPM configurations. Fig. 3(b) illustrates the field intensity within the PMs, which is well below the coercivity of the magnets (~870 kA/m), ensuring a high demagnetization tolerance. Fig. 3(c) shows the flux density distribution on the surface of iron yokes, PMs, and the CS. To achieve the smallest size, the outer yoke thickness is designed so that its maximum flux density, at the point where a flux loop closes its path, is close to the saturation point of the laminations. Fig. 3(d) illustrates one loop of the induced eddy currents in the CS and its magnitude distribution. The current density vectors close their path within the overhang regions. The interaction of the currents and magnetic field produces a Lorentz Force. As depicted in Fig. 2, since the reaction fields increase the flux density of one pole and decrease the flux density of the other pole, the current density distribution is asymmetrical—higher behind one pole. Fig. 4 shows the prototyped device and the experimental setup. The torque-speed characteristic of the coupler is given in Fig. 5, demonstrating a good correlation between FEM and experimental results. The larger the load torque, the higher the relative speed between the two sides. To avoid overheating, the maximum transmittable torque is at a speed where the average current density reaches around 40-50 $A/mm^2$, depending on the thermal model of the device.

## IV. CONCLUSION

A novel IPM eddy-current coupler is proposed, designed, and prototyped. Magnetic fields and eddy currents are derived and analyzed using FEM. Experimental results confirm the design and simulation results. The PMs are not exposed to the reaction fields and heat generated by the eddy currents, and field intensity within the PMs stays in a safe range, which ensures higher demagnetization tolerance and thus a longer lifetime, rendering it well-suited for applications with limited accessibility like offshore wind generation. As illustrated in Fig. 6(a), one application of the eddy-current coupler is as an intermediate torque transmission path between the turbine and the generator, filtering out torque transients such as wind gusts and tower shadowing effects in the turbine, which helps stabilize the generator. These couplers have also been utilized in slip synchronous PM wind generators (SS-PMG), integrating the principles of both induction and synchronous generators. This eliminates the need for a gearbox and power electronic converter (lower cost), offers simple maintenance, and eliminates torque ripple compared to conventional configurations utilizing a wound rotor [4]. A disadvantage of IPM topologies is the inherently large flux leakages at the unused sides of the IPM structure. To overcome this, we intend to investigate the proposed topology given in Fig. 6(b), which integrates the PM sides of the coupler and the generator, utilizing both the inner and outer radius surfaces of the IPM structure.

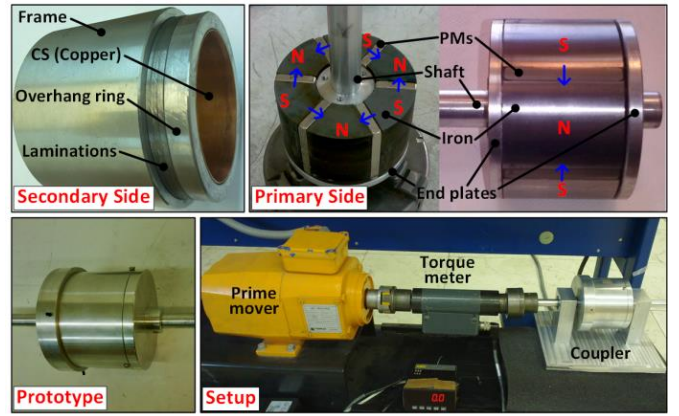

Fig. 4. Prototyped coupler and experimental setup

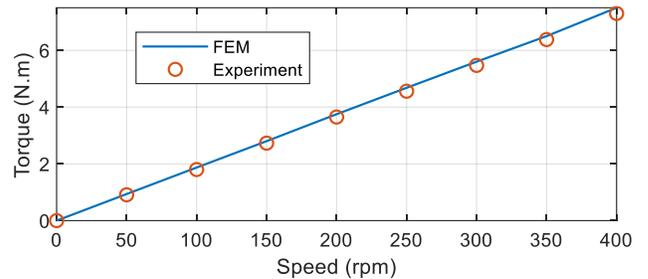

Fig. 5. Torque-speed characteristics of the proposed coupler

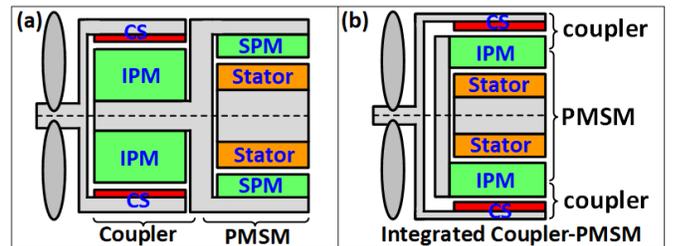

Fig. 6. Utilization of eddy-current couplers in wind turbines.